\documentclass[preprint,3p]{elsarticle} 
\usepackage{amssymb}
\usepackage{amsfonts}
\usepackage{amsmath}
\usepackage{amsthm}
\usepackage{algorithmic}
\usepackage{algorithm}
\usepackage{booktabs}
\usepackage{url}
\usepackage{multirow}
\usepackage{IEEEtrantools}
\usepackage{graphicx,color}
\usepackage[bf,SL,BF]{subfigure}
\usepackage{natbib}\biboptions{authoryear}
\usepackage{color,xcolor}
\usepackage[colorlinks, citecolor=blue, linkcolor=black]{hyperref}
\usepackage{enumerate}
\usepackage[title]{appendix}
\usepackage{lscape}
\usepackage{makecell}
\usepackage{colortbl}

\newcommand{\bx}{\mathbf{x}}\newcommand{\by}{\mathbf{y}} \def\b1{{\mathbf1}}

\newcommand{\bbR}{\mathbb{R}}\newcommand{\bbE}{\mathbb{E}}

\usepackage{xspace}
\newcommand{\et}{\emph{t}\xspace}

\newcommand{\tmeta}{\emph{t}Meta\xspace}
\newcommand{\tre}{\emph{t}RE-Meta\xspace}
\newcommand{\nmeta}{nMeta\xspace}
\newcommand{\mix}{MIX-Meta\xspace}
\newcommand{\symg}{SYM-Meta\xspace}
\newcommand{\skmg}{SKM-Meta\xspace}

\newcommand{\tsig}{\tilde{\sigma}^2}\newcommand{\tnu}{\widetilde{\nu}}

\newcommand{\cL}{\mathcal{L}}\newcommand{\cN}{\mathcal{N}}

\newcommand{\bbeta}{\boldsymbol{\beta}}

\newcommand{\btau}{\boldsymbol{\tau}}
\newcommand{\cQ} {\mathcal{Q}}
\newcommand{\ttau}{\tilde{\tau}}

\newcommand{\tmu}{\tilde{\mu}}
\newcommand{\tdelta}{\tilde{\delta}}
\newcommand{\hmu}{\hat{\mu}}
\newcommand{\hsig}{\hat{\sigma}^2}
\newcommand{\hnu}{\hat{\nu}}
\def\Beta{\mbox{Beta}}

 \newcommand{\bttheta}{\tilde{\btheta}}

\newcommand{\btheta}{\boldsymbol{\theta}}

\newcommand\refe[1]{(\ref{#1})}
\newcommand\reff[1]{Fig.~\ref{#1}}\newcommand\reft[1]{Table~\ref{#1}}\newcommand\refs[1]{Sec.~\ref{#1}} 
\newcommand\refa[1]{Algorithm~\ref{#1}}  
 
\theoremstyle{plain}  

\def\s2{\sigma^2}

\begin{document}	
		\title{A novel robust meta-analysis model using the \et distribution for outlier accommodation and detection}
  \begin{frontmatter}
        \author[J. Zhao]{Yue Wang}
 \author[J. Zhao]{Jianhua Zhao\corref{cor1}}\cortext[cor1]{Corresponding author.} \ead{jhzhao.ynu@gmail.com}
 \author[J. Zhao]{Fen Jiang}
 \author[J. Zhao]{Lei Shi}
 \author[J. Pan]{Jianxin Pan}

 \address[J. Zhao]{School of Statistics and Mathematics, Yunnan University of Finance and Economics, Kunming, 650221, China}
 \address[J. Pan]{Guangdong Provincial Key Laboratory of
Interdisciplinary Research and Application for
Data Science, BNU-HKBU United International
College, Zhuhai, 519087, China}
	
	\begin{abstract}
		Random effects meta-analysis model is an important tool for integrating results from multiple independent studies. However, the standard model is based on the assumption of normal distributions for both random effects and within-study errors, making it susceptible to outlying studies. Although robust modeling using the \et distribution is an appealing idea, the existing work, that explores the use of the \et distribution only for random effects, involves complicated numerical integration and numerical optimization. In this paper, a novel robust meta-analysis model using the \et distribution is proposed (\tmeta). The novelty is that the marginal distribution of the effect size in \tmeta follows the \et distribution, enabling that \tmeta can simultaneously accommodate and detect outlying studies in a simple and adaptive manner. A simple and fast EM-type algorithm is developed for maximum likelihood estimation. Due to the mathematical tractability of the \et distribution, \tmeta frees from numerical integration and allows for efficient optimization. Experiments on real data demonstrate that \tmeta is compared favorably with related competitors in situations involving mild outliers. Moreover, in the presence of gross outliers, while related competitors may fail, \tmeta continues to perform consistently and robustly.
	\end{abstract}
	
	
	\end{frontmatter}
	\section{Introduction}\label{sec:intr} 
	In meta-analyses, the collected studies often exhibit heterogeneity, characterized by greater variation among studies than can be explained by the variation within each study \citep{beath2014finite}, which could result in misleading conclusions about the overall treatment effect \citep{lin2017alternative,noma2022meta}. The random effects model is a popular tool for handling heterogeneity \citep{hardy1998detecting,wang2022penalization}. However, the standard model assumes normal distributions for both random effects and within-study errors (\nmeta), making it susceptible to outlying studies.

Outlier detection is a central research area in meta-analysis. Many methods have been developed. For example, a likelihood ratio test is constructed to identify outliers by detecting inflated variance \citep{gumedze2011random}; a forward search algorithm is developed specifically for this purpose \citep{mavridis2017detecting}; several outlier and influence diagnostic procedures in meta-regression models are presented \citep{viechtbauer2010outlier}. Subsequently, case deletion diagnostics and local influence analysis using multiple perturbation schemes, are investigated \citep{shi2017influence}. Several Bayesian outlier detection measures are also introduced for handling outlying studies in network meta-analysis \citep{zhang2015detecting}. Another important methodology for dealing with outliers is outlier accommodation or robust estimation, which can down-weight the influence of outliers. For instance, robust functions like Huber's rho and Tukey's biweight functions are employed to replace the original non-robust objective function, resulting in robust estimates \citep{yu2019robust}.

This paper focuses on outlier accommodation and detection simultaneously. Several efforts have been made toward this objective. Non-normal alternatives to normal random effects are investigated, and it is found that the \et distribution performs the best (\tre) \citep{baker2008new}. The shortcoming is that the marginal distribution of $y_i$ in \tre is mathematically intractable. Consequently, numerical integration is required to evaluate the log-likelihood and numerical optimization methods have to be employed for maximum likelihood (ML) estimation. Subsequently, new models where $y_i$ has a tractable marginal distribution are presented, including the three parameter symmetric marginal model (\symg) and the four parameter skew marginal model (\skmg) \citep{baker2016new}. Nevertheless, numerical optimization has still to be employed to obtain ML estimates. As a tractable model, a variant of a two-component mixture model (\mix) is proposed, with one component modeling standard studies and the other addressing outlying studies \citep{beath2014finite}. In \mix, the marginal distribution of the observed effect $y_i$ is a mixture of two normal distributions. However, \mix suffers from initialization issues, necessitating multiple runs of the fitting algorithm with different starting values. 
	
The common feature of these methods is that the error terms are assumed to follow the normal distribution. In this paper, we break this limitation as the marginal distribution of error term in our proposed model follows the \et distribution. It is known that the \et distribution includes the normal distribution as a special case when the degrees of freedom $\nu$ goes to infinity. This means that \tmeta offers greater flexibility and applicability than the conventional normal assumption. The main contributions of this paper are as follows. 
	
\begin{enumerate}[(i)]
\item The marginal distribution of the effect size $y_i$ in \tmeta follows the \et distribution, enabling it to simultaneously accommodate and detect outliers in a simple and adaptive manner. 1) The \et distribution offers an additional robustness tuning parameter which can adaptively down-weight outlying studies. 2) The expected weights follow in proportion to a Beta distribution, providing a useful critical value for outlier detection.
\item \tmeta provides a simple but powerful robust meta-analysis tool that can accommodate and detect both mild and gross outliers simultaneously. As can be seen from \refs{sec:re}, 1) \tmeta vs. \symg and \skmg. Both the three-parameter \symg and four-parameter \skmg fail in most of the outlier detection tasks, though they have good performance in outlier accommodation. 2) \tmeta vs. \tre and \mix. While all the three methods can be used to detect mild outliers, \tmeta performs the best in outlier accommodation. More importantly, in the presence of gross outliers, both \tre and \mix could fail while \tmeta still performs satisfactorily.
 
\item Due to its mathematical tractability, \tmeta frees from numerical integration and allows for efficient optimization. In contrast, \tre requires both complicated numerical integration and numerical optimization; \symg and \skmg involve complex numerical optimization \citep{baker2016new}; \mix requires multiple runs of the fitting algorithm due to the sensitivity issue of mixture models to initialization \citep{beath2014finite}. To our knowledge, \tmeta offers the first neat solution to robust meta-analysis modeling using the \et distribution.		 
\end{enumerate}

The rest of this paper is organized as follows. \refs{sec:back} reviews some related works. \refs{sec:tmeta} proposes our new model \tmeta. \refs{sec:re} conducts case studies to compare \tmeta with several closely related competitors. \refs{sec:Conclusion} offers a summary of the entire paper.  
	
	\section{Background}\label{sec:back}
In this section, we briefly review some fundamental results concerning the standard model \nmeta and Student's \et distribution.  
\subsection{Normal meta-analysis model (\nmeta)}\label{sec:nmeta}
In \nmeta, the effect size $y_i$ for the $i$-th study is defined as follows 
\begin{IEEEeqnarray}{rCl}\label{eqn:nme}
	y_i=\mu+b_i+e_i,\;i=1,\ldots,N,
\end{IEEEeqnarray}
where the random effects $b_i$ captures heterogeneity across studies and follows $\cN(0,\s2)$, the within-study error $e_i$ follows $\mathcal{N}(0, s_i^2)$ and they are independent of each other. Here, $\mu$ is the overall effect size, $\s2$ is the unknown between-study variance and $s_i^2$ is the known within-study variance.

From \refe{eqn:nme}, we have $y_i\sim\cN(\mu,\s2+s^2_i)$. Estimates for the parameters $\mu$ and $\sigma^2$ can be obtained through maximum likelihood methods \citep{hardy1996likelihood}.

\subsection{Student's \et distribution}\label{sec:t-dist}
Suppose that a random variable $y$ follows the univariate \et distribution $\et(\mu,\s2,\nu)$, with center $\mu\in\bbR$, scale parameter $\s2\in\bbR^{+}$, and degrees of freedom $\nu>0$, then the probability density function (p.d.f.) of $y$ is given by
\begin{IEEEeqnarray}{rCl}\label{eqn:t.pdf} 
	\nonumber f(y;\mu,\s2,\nu)=\frac{{\sigma^{-1}\Gamma(\frac{\nu+1}2)}}{{(\pi\nu)^{\frac12}}{\Gamma(\frac\nu2)}}{\left\lbrace 1+\frac{\delta^2(\mu,\s2)}{\nu}\right\rbrace^{-\frac{(\nu+1)}2}},
\end{IEEEeqnarray}
where $\Gamma(\cdot)$ is the gamma function and $\delta^2(\mu,\s2)=(y-\mu)^2/\s2$ is the squared Mahalanobis distance of $y$ from the center $\mu$ with respect to $\s2$. If $\nu>1$, $\bbE[y]=\mu$; if $\nu>2$, Var$(y)=\nu\s2/(\nu-2)$; and if $\nu\to\infty$, $t(\mu,\s2,\nu)\to\cN(\mu,\s2)$ \citep{liu1995ml}.

Given a latent weight variable $\tau$ distributed as the Gamma distribution $\mathrm {Gam}(\nu/2,\nu/2)$, $y$ can also be represented hierarchically as a latent variable model \citep{liu1995ml} as follows: 
\begin{IEEEeqnarray}{rCl}\label{eqn:t.latent}
	{y}|\tau\sim\cN(\mu,\frac{\s2}{\tau}),\,\,\tau\sim \mathrm{Gam}(\frac{\nu}2,\frac{\nu}2).
\end{IEEEeqnarray}
Under model \refe{eqn:t.latent}, it is easy to obtain the marginal distribution $y\sim \et(\mu,\s2,\nu)$ by $f(y;\mu,\s2,\nu)=\int_{0}^{\infty}f(y|\tau)f(\tau)d\tau$ \citep{zhao2006-rpca-t} and the posterior distribution of $\tau$ given $y$
\begin{IEEEeqnarray*}{rCl}\label{eqn:tau.post}
	\tau|y\,\,\sim \mathrm{Gam}\left(\frac{\nu+1}{2},\frac{\nu+\delta^2(\mu,\s2)}{2}\right).
\end{IEEEeqnarray*}
	  
\section{Novel robust meta-analysis model}\label{sec:tmeta}
In this section, we propose a novel robust meta-analysis model called \tmeta. In \refs{sec:tmeta.model}, we present the model. In \refs{sec:tmeta.mle}, we develop an algorithm for parameter estimation. In \refs{sec:outa} and \refs{sec:outd}, we give the details for outlier accommodation and detection in \tmeta.

\subsection{The proposed \tmeta model }\label{sec:tmeta.model}
Based on the hierarchical representation of the \et distribution in \refs{sec:t-dist}, we propose a novel robust random effects meta-analysis model, denoted by \tmeta. Its latent variable model can be expressed by
\begin{IEEEeqnarray}{rCl}\label{eqn:tme}
	\begin{cases}
		y_i=\mu+b_i+e_i,& i=1,\ldots,N,\\ 
		b_i|\tau_i\sim\cN(0,\s2/\tau_i), & 
		e_i|\tau_i\sim\cN(0,s^2_i/\tau_i)\\
		\tau_i\sim \mathrm{Gam}\left(\nu/2,\nu/2\right),&
	\end{cases}
\end{IEEEeqnarray}
where given the latent weight $\tau_i$, the random effects $b_i$ and the within-study error $e_i$ are mutually independent; $\mu$ is the overall effect size, $\s2$ is the unknown between-study variance, $s_i^2$ is the known within-study variance, and the degrees of freedom $\nu>0$.

Using the property of the normal distribution, it is easy to obtain the conditional distribution of $y_i$ given $\tau_i$
\begin{equation*}
y_i|\tau_i\sim\cN\left(\mu,\frac1{\tau_i}(\s2+s^2_i)\right),
\end{equation*}
and hence the marginal distribution 
\begin{equation*}
	y_i\sim t(\mu,\s2+s^2_i,\nu).
\end{equation*}
Moreover, it is known that the $t$ distribution $t(\mu,\s2+s^2_i,\nu)$ approaches the normal distribution $\cN(\mu,\s2+s^2_i)$ as $\nu\to\infty$, and thus \nmeta emerges as a special case of \tmeta in the limiting case.

\subsubsection{Probability distributions}\label{sec:p-dist}
From \tmeta model \refe{eqn:tme}, it is easy to obtain the following probability distributions
\begin{IEEEeqnarray}{rCl}
	 y_i|b_i,\tau_i&\sim&\cN\left(\mu+b_i,\frac{s^2_i}{\tau_i}\right),\,\,b_i|\tau_i\sim\cN\left(0,\frac{\s2}{\tau_i}\right),\nonumber \\
	 b_i|y_i,\tau_i&\sim&\cN\left(\frac{\s2(y_i-\mu)}{\s2+s^2_i}, \frac{\s2 s^2_i}{\tau_i(\s2+s^2_i)}\right),\nonumber\\
	 b_i|y_i&\sim& t\left(\frac{\s2(y_i-\mu)}{\s2+s^2_i}, \frac{\s2 s^2_i}{(\s2+s^2_i)},\nu\right),\nonumber\\
	\tau_i|y_i&\sim&\mathrm{Gam}\left(\frac{\nu+1}2,\frac{\nu+\delta^{2}_i(\mu,\sigma^{2})}2\right),\label{eqn:tau.y}
\end{IEEEeqnarray}
where 
\begin{equation}\label{eqn:mdist}
	\delta_i^2(\mu,\s2)=\frac{(y_i-\mu)^2}{\s2+s^2_i},
\end{equation}
is the squared Mahalanobis distance of $y_i$ from the overall effect size $\mu$. It is evident that all the probability distributions under \tmeta are well-known and tractable.

\subsubsection{Robust meta-regression with covariates}
When several covariates are involved, the model \refe{eqn:tme} can be extended to a more general model,
\begin{eqnarray*}\label{eqn:tme-rg}
	y_i=\bx_i'{\bbeta}+b_i+e_i,\quad i=1,\ldots,N, 
\end{eqnarray*}
where $\bx_i$ represents $p$-dimensional vector of covariates, $\bbeta=(\beta_1,\beta_2,\ldots,\beta_p)'$ is the $p$-dimensional regression coefficients; the random variables $b_i$ and $e_i$ and the other parameters $\mu, \s2$, and $\nu$ are similar as those in \tmeta \refe{eqn:tme}. Under this model, we have ${y_i}\sim t(\bx_i'{\bbeta},\sigma^{2}+s^{2}_i,\nu)$. 

\subsection{Maximum likelihood estimation}\label{sec:tmeta.mle}
In this section, we develop estimation algorithms for obtaining the ML estimates of the parameters $\btheta=(\mu,\s2,\nu)$ in the \tmeta model. Given the effect size vector $\by=(y_1,\ldots,y_N)$, from \refe{eqn:tme} the observed data log-likelihood function $\cL$ is (up to a constant),
\begin{IEEEeqnarray}{rCl} \label{eqn:tme.L}
	\cL(\btheta|\by)&=&-\>\frac12\sum\nolimits_{i=1}^N\left\lbrace(\nu+1)(\nu+\delta^2_i(\mu,\s2))+\ln(\s2+s^2_i)\right\rbrace \nonumber\\
	&&+\>N\left\lbrace\ln\Gamma(\frac{\nu+1}{2})-\ln\Gamma(\frac\nu2)+\frac{\nu}2\ln{\nu}\right\rbrace.
\end{IEEEeqnarray}

To maximize $\cL$ in \refe{eqn:tme.L}, we shall use an EM-type algorithm because of its simplicity and stability \citep{liu1995ml}. Specifically, we use an Expectation Conditional Maximization of Either (ECME) algorithm, a variant of the EM algorithm with faster monotone convergence \citep{liu1995ml}. Our ECME consists of an E-step followed by three conditional maximization (CM)-steps. In each CM step, a parameter in $\btheta=(\mu,\s2,\nu)$ is maximized while keeping the others fixed. 

Let the missing data be $\btau=(\tau_i,\ldots,\tau_N)$. From \refe{eqn:tme}, the log-likelihood function of complete data $(\by,\btau)$ is given by $$\cL_c(\btheta|\by,\btau)=\sum\nolimits_{i=1}^N\ln\{p(y_i|\tau_i)p(\tau_i)\}.$$ 
\textbf{E-step:} Compute the expected complete data log-likelihood function $\cL_c$ with respect to the conditional distribution $p(\boldsymbol{\tau}|\by,\btheta)$,
\begin{IEEEeqnarray*}{rCl}    
 \cQ(\btheta)=\bbE[\cL_{c}(\btheta|\by,\btau)|\by]=\cQ_1(\mu,\s2)+\cQ_2(\nu),
\end{IEEEeqnarray*}
where, up to a constant
\begin{IEEEeqnarray}{rCl}\label{eqn:tme.Q1}    
\hskip-1.5em \cQ_1(\mu,\s2)&=&-\frac12\sum\nolimits_{i=1}^N\left\{\ln(\s2+s^{2}_i)+\bbE[\tau_i|y_i]{\delta^{2}_i(\mu,\sigma^{2})}\right\}.
\end{IEEEeqnarray}
Here, $\delta^{2}_i(\mu,\sigma^{2})$ is given by \refe{eqn:mdist}.

From \refe{eqn:tau.y}, the required conditional expectation can be obtained as
\begin{IEEEeqnarray}{rCl}\label{eqn:Etau}
	\ttau_i\triangleq\bbE[\tau_i|y_i]=\frac{\nu+1}{\nu+{\delta_i^{2}}(\mu,\sigma^{2})},
\end{IEEEeqnarray}

In our ECME, the first two CM-steps maximize $\cQ$ while the third CM-step maximize $\cL$. In detail,

\noindent\textbf{CM-step 1:} Given $(\s2, \nu)$, maximize $\cQ_1$ in \refe{eqn:tme.Q1} with respect to $\mu$ yielding
\begin{equation}\label{eqn:tme.mu} 
	\left.\tmu=\sum\nolimits_{i=1}^N\frac{\ttau_i y_i}{\s2+s^2_i}\middle/\sum\nolimits_{i=1}^N\frac{\ttau_i}{\s2+s^2_i}\right..
\end{equation}
\textbf{CM-step 2:} Given $(\tmu, \nu)$, maximize $\cQ_1$ in \refe{eqn:tme.Q1} with respect to $\s2$ under the same restriction $\tsig\geq0$ as in \nmeta \citep{shi2017influence}, yielding
\begin{IEEEeqnarray}{rCl}
	\s2_t&=&\left.\sum\nolimits_{i=1}^N\frac{\ttau_i(y_i-\tmu)^2-s^2_i}{(\s2+s^2_i)^2}\middle/\sum\nolimits_{i=1}^N\frac{1}{(\s2+s^2_i)^2}\right.,\nonumber\\
	\tsig&=&\max\left\{\s2_t,\,0 \right\}.\label{eqn:tmeta.s2}
\end{IEEEeqnarray}

\noindent\textbf{CM-step 3:} Given $(\tmu, \tsig)$, maximize the observed data log-likelihood function $\cL$ in \refe{eqn:tme.L} w.r.t. $\nu$. This is equivalent to finding the root of the following equation
\begin{IEEEeqnarray}{rCl}\label{eqn:tme.nu}
	\cL'(\nu)&=&
	-\>\psi(\frac{\nu}2)+\ln(\frac{\nu}2)+1+\psi(\frac{\nu+1}{2})-\ln(\frac{\nu+1}{2})\nonumber\\
	&&\hskip-1.7em +\>\frac{1}{N}\sum\nolimits_{i=1}^N\left\lbrace\ln\left( \frac{\nu+1}{\nu+{\tdelta^2_i}}\right)-\left(\frac{\nu+1}{\nu+{\tdelta^2_i}}\right) \right\rbrace=0,
\end{IEEEeqnarray}
where $\tdelta^{2}_i=\delta_i^2(\tmu,\tsig)$, and $\psi(x)={d\ln(\Gamma(x))}/{dx}$ is the digamma function. Solving \refe{eqn:tme.nu} only requires one-dimensional search, which can be performed, e.g., by the bisection method \citep{liu1995ml}.

For clarity, the complete ECME algorithm is summarized in \refa{alg:ecme}.
\begin{algorithm}[htb]
	\caption{The ECME algotithm for \tmeta}
	\label{alg:ecme}
	\begin{algorithmic}[1]
		\REQUIRE Data $\by$ and initialization of $\btheta$=($\mu,\s2,\nu$).
		\REPEAT
		\STATE \emph{E-step:} Compute $\ttau_i$ via \refe{eqn:Etau}.
		\STATE \emph{CM-step 1:} Update $\tmu$ via \refe{eqn:tme.mu}.
		\STATE \emph{CM-step 2:} Update $\tsig$ via \refe{eqn:tmeta.s2}. 	
		\STATE \emph{CM-step 3:} Update $\tnu$ via \refe{eqn:tme.nu}.		
		\UNTIL\textbf{until} the relative change of $\cL$ in \refe{eqn:tme.L} is smaller than a threshold.	
		\ENSURE $\bttheta=(\tmu,\tsig,\tnu)$.
	\end{algorithmic}
\end{algorithm}

\subsection{Outlier accommodation}\label{sec:outa}
\subsubsection{Adaptive outlier accommodation}
Looking at \refe{eqn:Etau}, \refe{eqn:tme.mu} and \refe{eqn:tmeta.s2}, the following can be observed.
\begin{enumerate}[(i)]
	\item When the data contain no outliers and the $y_i$'s come from \nmeta, $\nu$ is expected to take on large values. This causes all the weights $\ttau_i$ in \refe{eqn:Etau} to be close to 1. Consequently, \refe{eqn:tme.mu} and \refe{eqn:tmeta.s2} would degenerate to those of \nmeta, and hence \tmeta adaptively degenerates to \nmeta in this case.
	\item In the presence of outliers, $\nu$ is expected to take on small values, and the outlying study $y_i$ would have a much greater squared Mahalanobis distance $\delta^{2}_i(\mu,\s2)$ compared with non-outliers, causing the outlier's $\ttau_i$ in \refe{eqn:Etau} to be much smaller than those of non-outliers. Consequently, the impact of outliers on the estimators in \refe{eqn:tme.mu} and \refe{eqn:tmeta.s2} is substantially reduced, allowing \tmeta to yield robust estimates.
\end{enumerate}
In summary, the degrees of freedom $\nu$ is a robustness tuning parameter that adapts according to the presence of outliers in the data.

\subsubsection{Breakdown point}
In statistics, the robustness of estimators is assessed by breakdown points, which are the proportion of arbitrarily large outlying observations an estimator can tolerate before giving an incorrect result. The following Proposition 1 gives the breakdown point of \tmeta.\\
\textbf{Proposition 1}.The upper bound of the breakdown point of \tmeta is $1/(\nu+1)$.
\begin{proof}
As proved by 
\cite{dumbgen2005breakdown},  the upper bound of the breakdown point of the $d$-dimensional multivariate \emph{t} distribution is $1/(\nu+d)$. For \tmeta, the dimension of \et-distributed $y_i$ is $d=1$ and hence the upper bound of \tmeta is given by $1/(\nu+1)$. This completes the proof.
\end{proof}
In our implementation, we restrict $\nu\geq 1$. Proposition 1 shows that \tmeta is a highly robust method as its breakdown point could be close to 50\% under this restriction. 

\subsection{Outlier detection}\label{sec:outd}
Similar to that in multivariate \et and matrix-variate \et distributions \citep{wang2011,zhao2023-rfpca,zhao2023-tbppca}, the expected weight $\ttau_i$ in \tmeta given by \refe{eqn:Etau} can be used as outlier indicator. Let \begin{IEEEeqnarray}{rCl}\label{eqn:ui}
	\left. u_i=\frac{N}{\hsig+s^2_i}\middle/\sum\nolimits_{i=1}^N\frac1{\hsig+s^2_i}\right..
\end{IEEEeqnarray}
The following Proposition 2 gives the details.\\
\textbf{Proposition 2}. Assume that the study $\{y_i\}_{i=1}^N$ follow \tmeta model \refe{eqn:tme}. Given the ML estimate $\hat{\btheta}$, we have, when the estimate $\hsig>0$,
\begin{IEEEeqnarray*}{rCl}\label{eqn:m.Etau}
	\frac1N\sum\nolimits_{i=1}^Nu_i\ttau_i=1,
\end{IEEEeqnarray*}
and when $\hsig=0$,
\begin{IEEEeqnarray*}{rCl}\label{eqn:m.Etau0}
	\frac1N\sum\nolimits_{i=1}^Nu_i\ttau_i\geq1,
\end{IEEEeqnarray*}
\begin{proof}
	The proof can be found in \refs{m.Etau}.
\end{proof}
\noindent\textbf{Proposition 2} shows that when the estimate $\hsig>0$, the average of all $u_i\ttau_i$'s equals to 1. In other words, the study with $u_i\ttau_i$ much smaller than 1 (i.e., $\ttau_i$ much smaller than $1/u_i$) or close to 0 can be considered as an outlier. When $\hsig=0$, our experience reveals that $\sum\nolimits_{i=1}^Nu_i\ttau_i/N$ may be slightly greater than $1$. 

In practice, a critical value is needed to judge whether a study is an outlier or not. The following Proposition 3 does this task. Let $F(a,b)$ and $\Beta(a,b)$ stand for the $F$ distribution and Beta distribution with parameters $a$ and $b$, respectively. The $\alpha$ quantile of $\Beta(a,b)$ is denoted by $B_\alpha$. \\
\textbf{Proposition 3}. Suppose that the study set $\{y_i\}_{i=1}^N$ follow \tmeta model \refe{eqn:tme}. Then we have that the Mahalanobis distance $\delta^2_i(\mu,\s2)\sim F(1,\nu)$. Given the ML estimate $\hat{\btheta}$, the weights $\ttau_i,\,i=1,\ldots,N$ converge in distribution to $(1+1/\nu)Beta(\nu/2,1/2)$ as the study sample size $N$ approaches infinity. Therefore, at a significance level of $\alpha$, the $i$-th study with $\ttau_i<(1+1/\nu)Beta_{\alpha}(\nu/2,1/2)$ could be identified as an outlier.
\begin{proof}
This is a special case with dimension $d=1$ of the result on the $d$-dimensional multivariate \et distribution proved by \cite{wang2011}. This completes the proof.
\end{proof}

\section{Results}\label{sec:re} 
In this section, we compare the performance of our proposed \tmeta with five closely related methods: \nmeta, \tre, \mix, \symg and \skmg using four benchmark real-world datasets. For \tmeta, the iteration stops when the relative change in the objective function $\cL$ (|1-$\cL^{(t)}/\cL^{(t+1)}$|) is smaller than the given threshold $tol=10^{-8}$ or the number of iterations exceeds $t_{max}=100$. For \nmeta, \tre, and \mix, we use the R codes available from \url{https://cran.r-project.org/web/packages/metaplus/}. In addition, we use the default setting for \mix, i.e., 20 initializations. The code for \symg and \skmg can be found from the supplementary materials by \cite{baker2016new}.  

To perform outlier detection for \tmeta, we utilize the critical value provided in Proposition 3 and set the significance level $\alpha=0.05$. For better visualization, we equivalently plot the inverse of $\ttau_i$. That is, the study with $1/\ttau_i>1/((1+1/\nu)Beta_{\alpha}(\nu/2,1/2))$ is identified as an outlier for \tmeta. For \mix, we use the empirical threshold 0.9 as suggested by \cite{beath2014finite} which represents the posterior probability that a study belongs to the outlying component. For \symg and \skmg, we adopt the $p$-value method specially developed for both models by \cite{baker2016new} Since \tre lacks guidelines for setting the threshold, we follow the empirical approach by \cite{baker2008new} treating studies with very small values of the relative weight $\omega_i/\omega_i^0$, or equivalently, very large values of $\omega_i^0/\omega_i$ as outliers, where $\omega_i$ and $\omega_i^0$ are the weights under \tre and \nmeta, respectively.

To compare the computational efficiency, we will report their total CPU time consumed by various methods, which is sum of the time used for training model and that for detecting outliers. For \tmeta and \mix, outlier detection is a byproduct of the model training and incurs no additional time cost. However, \tre, \symg and \skmg require additional time cost for outlier detection. To be specific, \tre requires numerical methods to compute $\omega_i^0/\omega_i$ while \symg and \skmg necessitate additional efforts to implement the $p$-value method. 

\subsection{Intravenous magnesium}\label{sec:mag}
\begin{figure}[!t]
	\centering
	\scalebox{0.65}[0.65]{\includegraphics*{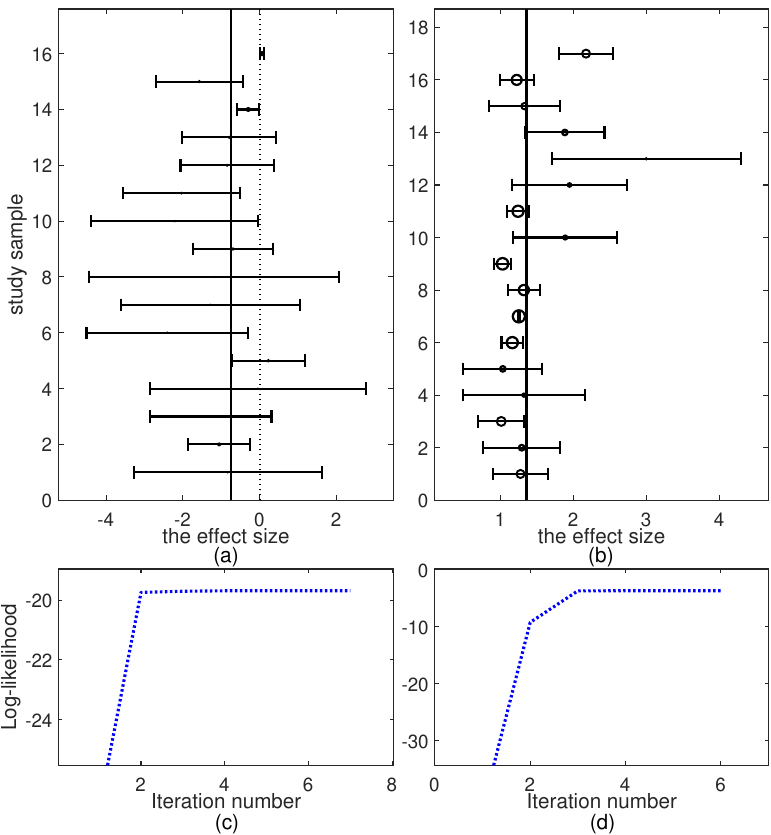}}
	\caption{Top row: forest plots on two datasets: (a) Mag and (b) Hipfrac, where each effect size $y_i$ and 95\% confidence interval are shown as circle and solid line, respectively. Bottom row: evolement of log-likelihood of $\cL$ versus number of iterations: (c) Mag and (d) Hipfrac.}
	\label{fig:mh.fore}
\end{figure}

The Mag dataset \citep{Sterne2001} comprises 16 studies. As can be seen from the forest plot shown in \reff{fig:mh.fore}~(a), it looks difficult to visually identify which study is an outlier except that study 16 seems different from others due to its relatively large $y_i$ value and low $s_i^2$. Previous researches \citep{gumedze2011random,beath2014finite} have analyzed this dataset and found no outliers. Below we perform outlier detection with various methods.

We fit all the six methods on the Mag dataset. \reft{tbl:mag} collects the results. The results in \reft{tbl:mag} show all the six methods yield similar performance. This means that all the five methods \tre, \mix, \symg, \skmg and \tmeta could degrade to \nmeta. Nevertheless, among the five robust methods, \tmeta is computationally the most efficient while \tre and \mix require much more time. \reff{fig:mh.fore}~(c) shows the evolvement of log-likelihood $\cL$ versus number of iterations when fitting \tmeta. It can be seen from \reff{fig:mh.fore}~(c) that \tmeta converges within 7 iterations on this dataset.

\begin{table*}[htbp]
	\centering
	\caption{\label{tbl:mag} Results of parameter estimates, negative log-likelihood, and CPU time by various methods on Mag dataset. The best method is shown in boldface. `---' indicates that a method does not have corresponding results. }
		\begin{tabular}{cccccc}  
			\toprule
			Methods &$\mu$ &$\sigma$ &$\nu$ &-$\cL$ &Time\\ \midrule
			\nmeta &-0.746 &0.504 &--- &19.685 &---\\ 
			\tre &-0.746 &0.504 &inf &19.685 &3.0\\ 
			\mix &-0.746 &0.504 &--- &19.685 &32.4\\
			\symg &-0.746 &0.504 &--- &19.685 &0.3\\
			\skmg &-0.746 &0.504 &--- &19.685 &0.3\\
			\tmeta &-0.746 &0.504 &inf &19.685 &0.05\\
			\bottomrule
		\end{tabular}
\end{table*}

\begin{figure*}[!t]
	\centering
	\scalebox{0.85}[0.85]{\includegraphics*{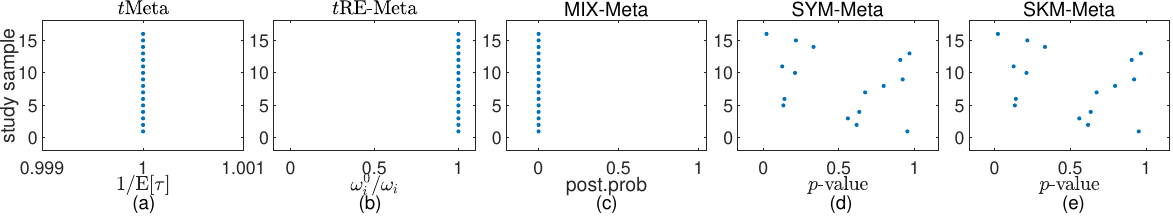}}
	\caption{Results on outlier detection by various methods on Mag dataset: (a) \tmeta; (b) \tre; (c) \mix; (d) \symg; (e) \skmg. The marker solid point $\bullet$ in blue represents normal studies judged by a method.}\label{fig:m.dec} 
\end{figure*}

\reff{fig:m.dec} shows the results of detecting outliers by the five methods. It can be seen from \reff{fig:m.dec} that all the five methods suggest no outliers for Mag dataset. This finding is consistent with that by \cite{beath2014finite}.

\subsection{Hipfrac dataset}
The Hipfrac dataset \citep{haentjens2010meta} contains 17 studies, collected from an investigation on the magnitude and duration of excess mortality after hip fracture among older men. \reff{fig:mh.fore}~(b) shows the forest plot, from which it seems hard to identify which study is an outlier. Below we perform outlier analysis with various methods.

We fit all the six methods on the Hipfrac dataset. \reft{tbl:hipfrac} summarizes the results. The results in \reft{tbl:hipfrac} show that \tmeta and \skmg obtain significantly better BIC than the other methods and \skmg wins by a narrow margin. In terms of computational efficiency among the five robust methods, \tmeta is the fastest while \tre and \mix are the slowest runners. \reff{fig:mh.fore}~(d) shows the evolution of log-likelihood $\cL$ versus number of iterations when fitting \tmeta. It can be seen from \reff{fig:mh.fore}~(d) that \tmeta converges within 6 iterations on this dataset.

\begin{table*}[!t]
	\centering
	\caption{\label{tbl:hipfrac} Results of parameter estimates, negative log-likelihood, BIC, and CPU time by various methods on Hipfrac dataset. `---' indicates that a method does not have corresponding results.}
		\begin{tabular}{ccccccc}
			\toprule
			Methods &$\mu$ &$\sigma$ &$\nu$ &-$\cL$ &BIC & Time\\ \midrule
			\nmeta &1.357 &0.260 &---&8.498  &22.661 &---\\ 
			\tre &1.251 &0.013 &0.582 &6.575  &21.649 &79.2\\ 
			\mix &1.252 &0.000 &--- &4.507  &20.347 &364.1\\
			\symg &1.220 &0.074 &--- &5.670  &19.840 &0.222\\
			\skmg &1.202 &0.063 &--- &1.439 &\textbf{14.212} &0.2\\
			\tmeta &1.252 &0.000 &1.871 &3.700 &15.899 &0.03\\
			\bottomrule
		\end{tabular}
\end{table*}

\begin{figure*}[!t]
	\centering
	\scalebox{0.85}[0.85]{\includegraphics*{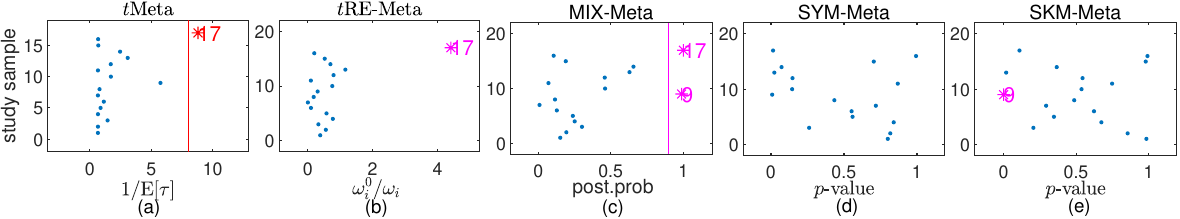}}
	\caption{Results on outlier detection by various methods on Hipfrac dataset: (a) \tmeta; (b) \tre; (c) \mix; (d) \symg; (e) \skmg. The vertical line indicates the critical value for \tmeta (red) and the threshold 0.9 (magenta) for \mix. The vertical line indicates the critical value for \tmeta and the threshold 0.9 for \mix. The marker solid point $\bullet$ in blue represents normal studies judged by a method. Star `*' signals outlying studies, with red for \tmeta and magenta for the other methods.} \label{fig:h.dec}
\end{figure*}

\reff{fig:h.dec} shows the results of detecting outliers by the five methods. It can be seen from \reff{fig:h.dec} that both \tmeta and \tre identify study 17 as an outlier. This result is consistent with that obtained by \cite{lin2017alternative}. In contrast, \mix identifies one more outlier: study 9, while \symg and \skmg fail completely. In fact, from \reff{fig:mh.fore}~(b), it seems not plausible to treat study 9 as an outlier.

\subsection{Fluoride toothpaste}\label{sec:flu}
This dataset contains 70 studies, obtained from an evaluation of fluoride's efficacy in preventing childhood dental caries \citep{marinho2003}. The effect size $y_i$ denotes the difference between control and treatment groups, with negative values signifying significant therapeutic effects. 

Previous works \citep{baker2008new,gumedze2011random,beath2014finite} have concluded that there exist three outliers in this dataset: study 63, study 50 and study 38. Contrarily, the analysis with \skmg suggests no outliers in the dataset \citep{baker2016new}. To better examine the outlier detection performance by various methods, we shall perform two experiments in this section. In the first experiment of \refs{sec:flu.org}, we use the original dataset (Flu). In the second experiment of \refs{sec:flu.ext}, we add the original dataset with one more artificial outlier. The resulting dataset is called modified Flu for clarity.

\subsubsection{Original Flu}\label{sec:flu.org}
\begin{figure}[!t]
	\centering
	\scalebox{0.65}[0.65]{\includegraphics*{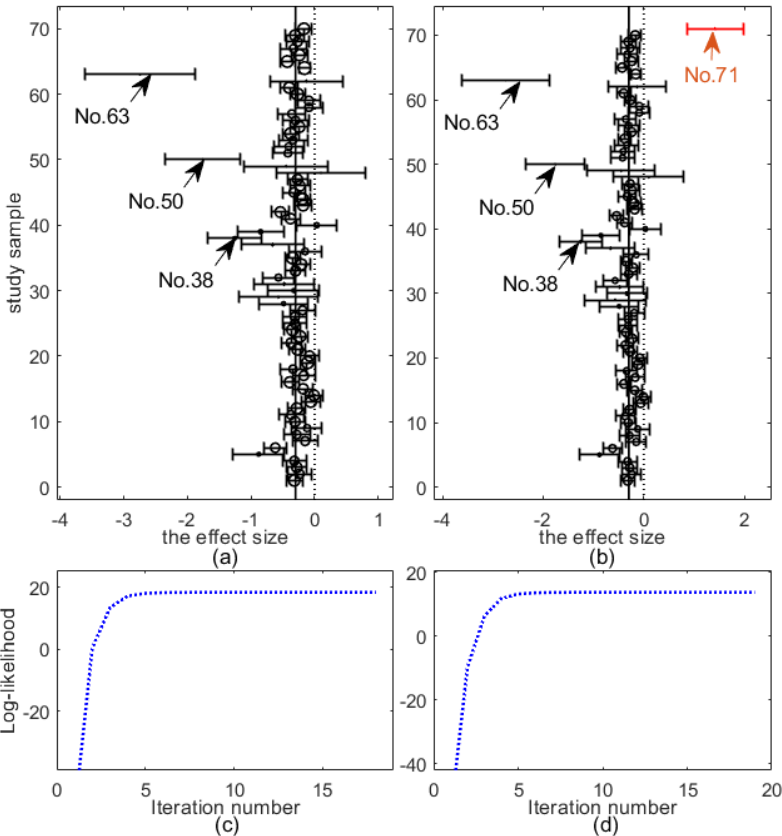}}
	\caption{Top row: forest plots on the fluoride toothpaste dataset: (a) Flu and (b) modified Flu, where each effect size $y_i$ and 95\% confidence interval are shown as circle and solid line, respectively. Bottom row: evolement of log-likelihood of $\cL$ versus number of iterations: (c) Flu and (d) modified Flu.} \label{fig:f.fore}
\end{figure}

\reff{fig:f.fore}(a) shows the forest plot of the original dataset Flu. It can be observed from \reff{fig:f.fore}(a) that studies 38, 50, and 63 look like abnormal. We then perform further analysis to identify outliers.

We fit all the six methods on Flu. \reft{tbl:Fluoride} summarizes the results. The results in \reft{tbl:Fluoride} show that \tmeta, \symg and \skmg obtain substantially better BIC than the other methods and \skmg is again the best. Among the five robust methods, \tmeta is the most computationally efficient while \tre and \mix are the most inefficient. \reff{fig:f.fore}~(c) shows the evolution of log-likelihood $\cL$ versus number of iterations when fitting \tmeta. It can be seen from \reff{fig:f.fore}~(c) that \tmeta converges within 18 iterations on this dataset.

\begin{table*}[!t]
	\centering
	\caption{\label{tbl:Fluoride} Results by various methods on the original and modified fluoride toothpaste dataset, including parameter estimates, negative log-likelihood, BIC, and CPU time. The best method is shown in boldface. `---' indicates that a method does not have corresponding results. }
	
	\begin{tabular}{ccccccc}
			\toprule
			Methods     &$\mu$  &$\sigma$ &$\nu$ &-$\cL$ &BIC &Time\\ \midrule
			&\multicolumn{6}{c}{Original Flu}\\
			\nmeta &-0.300 &0.119 &---     &1.233 &10.963 &---\\ 
			\tre &-0.280 &0.049 &1.158 &-13.121 &-13.497 &64.0\\ 
			\mix &-0.281 &0.090 &--- &-14.636 &-12.277 &27.0\\
			\symg &-0.282 &0.092 &--- &-17.148  &-21.551 &0.4\\
			\skmg &-0.273 &0.081    &--- &-21.914  &\textbf{-26.834} &0.6\\
			\tmeta &-0.282 &0.051 &2.754 &-18.283 &-23.820 &0.05\\
			\cmidrule{2-7} 
			& \multicolumn{6}{c}{Modified Flu}\\
			\nmeta &-0.297 &0.139 &--- &15.760 &40.046 &---\\
			\tre &-0.279 &0.047 &1.023 &-7.774  &-2.761 &59.2\\
			\mix &-0.280 &0.088 &--- &-10.062 &-3.072 &26.9\\
			\symg &-0.282 &0.092 &--- &-12.399 &-12.010 &0.7\\
			\skmg &-0.277 &0.088 &--- &-13.144 &-9.238 &0.6\\
			\tmeta &-0.281 &0.047 &2.367 &-13.791 &\textbf{-14.794} &0.05\\
			\bottomrule
		\end{tabular}
\end{table*}

\begin{figure*}[ht]
	\centering
	\scalebox{0.85}[0.85]{\includegraphics*{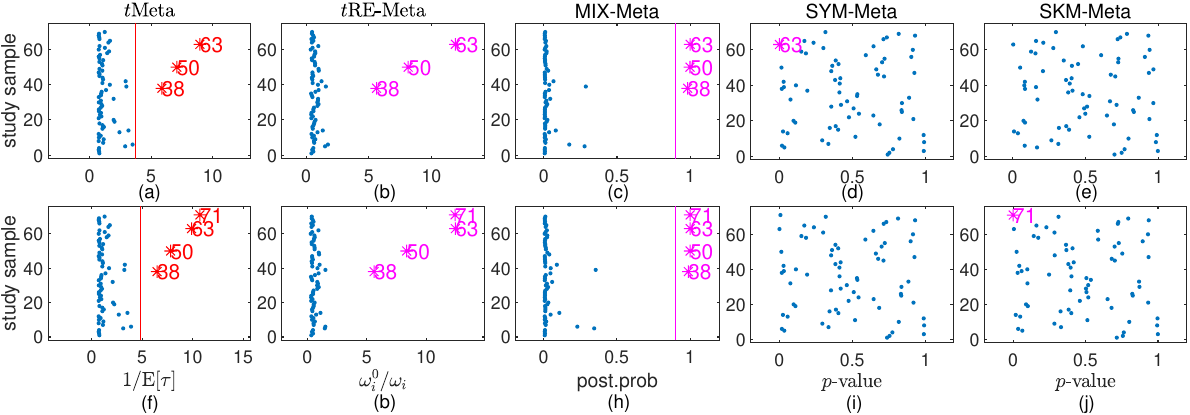}}
	\caption{Results on outlier detection by various methods on fluoride toothpaste dataset. Top row: the original dataset; Bottom row: the modified dataset. (a), (f) \tmeta; (b), (g) \tre; (c), (h) \mix; (d), (i) \symg; (e), (j) \skmg. The vertical line indicates the critical value for \tmeta and the threshold 0.9 for \mix. The marker solid point $\bullet$ in blue represents normal studies judged by a method. Star `*' signals outlying studies, with red for \tmeta and magenta for the other methods. } \label{fig:f.dec}
\end{figure*}

The top row in \reff{fig:f.dec} shows the results of detecting outliers by the five methods. It can be seen from \reff{fig:f.dec} that \tmeta, \tre and \mix all identify three studies: 63, 50, 38. This means that the result by \tmeta is consistent with those in previous works \citep{baker2008new,gumedze2011random,beath2014finite}. In contrast, \symg only detects the most abnormal study 63 as one outlier while \symg identify no outlier.

\subsubsection{Modified Flu}\label{sec:flu.ext}
In the modified Flu, the outlier (study 71) is introduced as follows. The effect size $y_{71}$ is generated from the
uniform distribution $U$ on the interval $[1, 2]$, i.e., $y_{71}\sim U(1,2)$. We set its within-study variance $s^2_{71}=1/12$. \reff{fig:f.fore}~(b) shows the forest plot of modified Flu, from which it can be seen that the newly added study 71 looks like a mild outlier as it is very different from all the other studies.

\reft{tbl:Fluoride} summarizes the results by six methods. The results in \reft{tbl:Fluoride} show that \tmeta yields the best BIC on this dataset, which is then followed by \symg, and \skmg is the third best. Among the five robust methods, \tmeta is again the best performer in computational efficiency while \tre and \mix are still the most inefficient. \reff{fig:f.fore}~(d) shows the evolution of log-likelihood $\cL$ versus number of iterations when fitting \tmeta. It can be seen from \reff{fig:f.fore}~(d) that \tmeta converges within 19 iterations on this dataset.

The bottom row in \reff{fig:f.dec} shows the results of detecting outliers by the five methods. It can be seen from the bottom row of \reff{fig:f.dec} that \tmeta, \tre and \mix successfully identify four outliers: 71, 63, 50, 38. In contrast, \symg fails to detect any outlier, while \skmg can detect the newly added study 71.  

\subsection{CDP-choline}\label{sec:cdp}
The CDP-choline dataset \citep{fioravanti2005} is obtained by exploring the cytidinediphosphocholine analysis in cognitive and behavioural disorders associated with chronic brain diseases in the elderly. The sample size is $N=10$. 

Previous analyses \citep{baker2008new,gumedze2011random,beath2014finite} have concluded that there is one outlier in this dataset: study 8. Like \refs{sec:flu}, we perform two experiments. In the first experiment of \refs{sec:cdp.org}, we use the original dataset (CDP). In the second experiment of \refs{sec:cdp.ext}, we modify CDP so that it contains more outliers, which is denoted by modified CDP for clarity. 

\subsubsection{Original CDP}\label{sec:cdp.org}
\begin{figure}[ht]
	\centering
	\scalebox{0.65}[0.65]{\includegraphics*{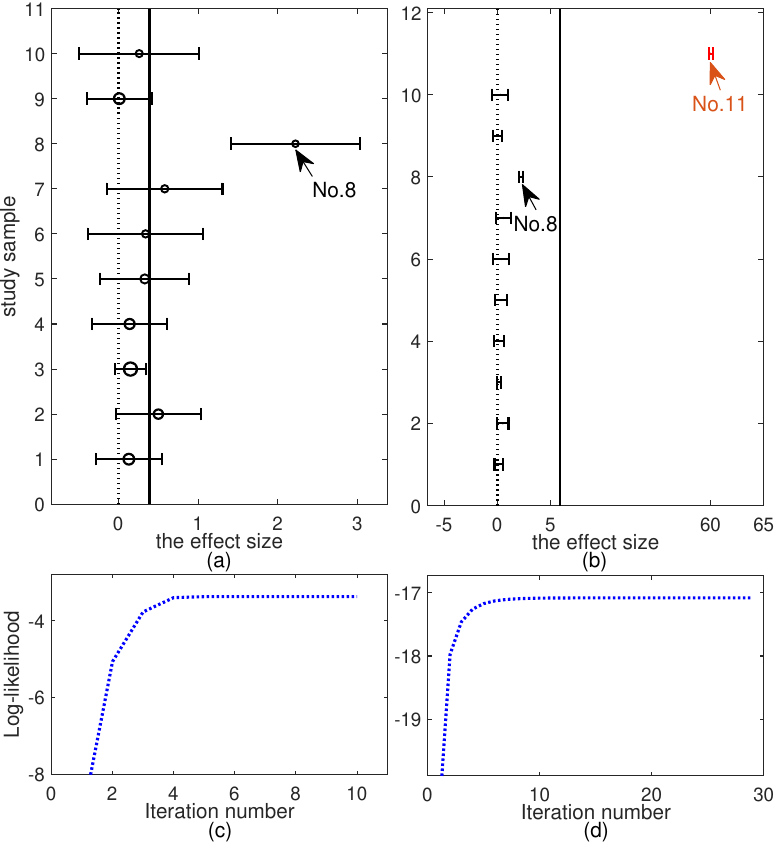}}
	\caption{Top row: forest plots on CDP-choline dataset: (a) original dataset; (b) modified dataset. Bottom row: evolement of log-likelihood of $\cL$ versus number of iterations: (c) original dataset and (d) modified dataset.}\label{fig:c.fore}
	\end{figure}

\reff{fig:c.fore}(a) shows the forest plot of the original CDP. It can be observed from \reff{fig:c.fore}(a) that study 8 looks like abnormal. We then perform further analysis to identify outliers.

We fit all the six methods on CDP. \reft{tbl:cdp} summarizes the results. The results in \reft{tbl:cdp} show that \tmeta, \symg and \skmg obtain significantly better BIC than the other methods and \skmg is again the best. In terms of computational efficiency among the five robust methods, \tmeta is the most efficient while \tre and \mix are the slowest runners. \reff{fig:c.fore}~(c) shows the evolution of log-likelihood $\cL$ versus number of iterations when fitting \tmeta. It can be seen from \reff{fig:c.fore}~(c) that \tmeta converges within 10 iterations on this dataset.
	
\begin{table*}[ht]
\centering
\caption{\label{tbl:cdp} Results by various methods on the CDP-choline dataset, including parameter estimates, negative log-likelihood, BIC, and CPU time. The best method is shown in boldface. `---' indicates that a method does not have corresponding results.}

	\begin{tabular}{ccccccc}
		 \toprule
		Methods &$\mu$ &$\sigma$ &$\nu$ &-$\cL$ &BIC &Time\\ \midrule
		&\multicolumn{6}{c}{Original CDP}\\
		\nmeta &0.389 &0.383 &--- &8.199 &21.002 &---\\ 
		\tre &0.195 &0.006 &0.494 &4.058 &15.024 &24.7\\ 
		\mix &0.191 &1.777 &--- &3.007 &15.225 &47\\
		\symg &0.194 &0.000 &--- &2.847  &12.602 &0.14\\
		\skmg &0.193 &0.000 &--- &1.403 &\textbf{12.016} &0.2\\
		\tmeta &0.187 &0.000 &2.380 &3.377 &13.662 &0.03\\ \cmidrule{2-7} 
		& \multicolumn{6}{c}{Modified CDP}\\
		\nmeta &5.879 &17.126 &--- &46.855 &98.506 &---\\
		\tre &0.193 &0.002 &0.273 &13.768 &34.729 &65.9\\
		\mix &5.879 &2.455 &--- &46.855 &103.302 &19.5\\
		\symg &5.880 &17.117 &--- &46.855  &100.904 &0.3\\
		\skmg &0.484 &0.711 &--- &21.622 &52.836 &0.4\\
		\tmeta &0.200 &0.115 &1.000 &17.081 &\textbf{41.355} &0.03\\
		\bottomrule
	\end{tabular}
\end{table*}

\begin{figure*}[ht]
\centering
\scalebox{0.85}[0.85]{\includegraphics*{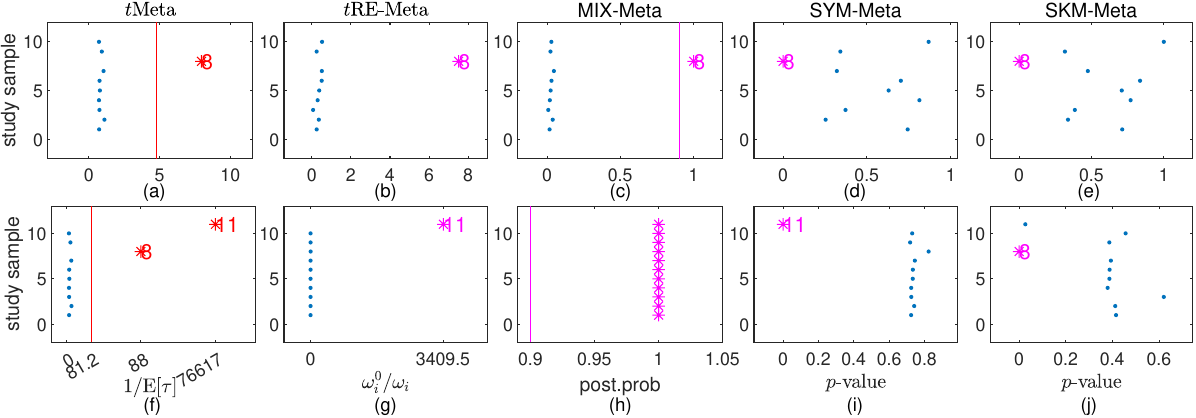}}
\caption{Results on outlier detection by various methods on CDP-choline dataset. Top row: the original dataset; Bottom row: the modified dataset. (a), (f) \tmeta; (b), (g) \tre; (c), (h) \mix; (d), (i) \symg; (e), (j) \skmg. The vertical line indicates the critical value for \tmeta and the threshold 0.9 for \mix. The marker solid point $\bullet$ in blue represents normal studies judged by a method. Star `*' signals outlying studies, with red for \tmeta and magenta for the other methods.}\label{fig:c.dec}
\end{figure*}

The top row in \reff{fig:c.dec} shows the results of detecting outliers by the five methods. It can be seen from \reff{fig:f.dec} that all the five methods successfully identify study 8 as an outlier. This means that the results by \tmeta, \symg and \skmg are consistent with those in previous works \citep{baker2008new,gumedze2011random,beath2014finite}.  

\subsubsection{Modified CDP}\label{sec:cdp.ext}
In the modified CDP, we make two modifications: (i) add one outlier, namely study 11, which is set as $y_{11}=60, s_{11}^2=0.25$; (ii) set $s_8^2=0.01$. \reff{fig:c.fore}~(b) shows the forest plot of modified CDP, from which it can be seen that the newly added study 11 is a gross outlier as it is extremely different from all the other studies and study 8 is a mild outlier but now it has a smaller within-study variance than that in the original CDP.

\reft{tbl:cdp} summarizes the results by six methods. The results in \reft{tbl:cdp} show that \tre and \tmeta have better BIC than the other methods on this dataset and \tre is the best. In terms of computational efficiency among the five robust methods, \tmeta is still the most efficient while \tre and \mix demand the most time. \reff{fig:c.fore}~(d) shows the evolution of log-likelihood $\cL$ versus number of iterations when fitting \tmeta. It can be seen from \reff{fig:c.fore}~(d) that \tmeta requires 29 iterations to converge on this dataset.

The bottom row in \reff{fig:c.dec} shows the results of detecting outliers by the five methods. It can be seen that \tmeta performs reliably as it successfully detects the two outliers: study 11, 8. In contrast, \mix and \skmg fail to detect the most extreme study 11. \tre and \symg can identify study 11 but they fail to detect study 8.

\section{Conclusion}\label{sec:Conclusion} 
For outlier accommodation and detection simultaneously, in this paper, we propose a novel robust meta-analysis model using student's \et distribution, namely \tmeta. \tmeta can be expressed as a hierarchical latent variable model while the marginal distribution of the effect size $y_i$ follows a tractable \et distribution. To obtain the ML estimates of the parameters, we develop an ECME algorithm, which is computationally much more efficient than related methods as shown in our experiments. Empirical results on real datasets show that \tmeta not only improves the robustness of \nmeta as expected but also is compared favorably with closely related competitors in that it can provide the best performance for outlier accommodation and detection simultaneously, for both mild and gross outliers. 

The experiment results show that \skmg on some datasets yields better performance in outlier accommodation. For future work it would be interesting to extend \tmeta using the skew-\et distribution for further accommodating skewed data. 
 
    \section*{Acknowledgments}
	This work was supported partly by the National Natural Science Foundation of China under Grant 12161089 and Grant 11931015; partly by Yunnan Province Xingdian Young Talent Support Program under Grant YNWR-QNBJ-2018-365.
	\begin{appendices}
	\section{Proof for Proposition 1}\label{m.Etau}
	\begin{proof}
		For ML estimate $\hat{\btheta}$, multiplying \refe{eqn:Etau} by $\hnu+\delta^2_i(\hmu,\s2)$, we obtain
		\begin{IEEEeqnarray}{rCl}\label{eqn:tme.Etau.t1}
		\hnu+1=\hnu\ttau_i+\ttau_i{\delta_i^{2}}(\hmu,\hsig).
		\end{IEEEeqnarray}
		On both sides of \refe{eqn:tmeta.s2}, multiply by $\sum\nolimits^N_{i=1}1/(\hsig+s^2_i)^2$ and then add $\sum\nolimits^N_{i=1}s^2_i/(\hsig+s^2_i)^2$. On noting \refe{eqn:mdist}, when $\hsig>0$, we have
		\begin{IEEEeqnarray}{rCl}\label{eqn:tmeta.s2.t} \sum\nolimits_{i=1}^N\frac{\ttau_i\delta^2_i(\hmu,\hsig)}{\hsig+s^2_i}=\sum\nolimits_{i=1}^N\frac1{\hsig+s^2_i},
		\end{IEEEeqnarray}
		and when $\hsig=0$, we have 
		\begin{IEEEeqnarray}{rCl}\label{eqn:tmeta.s2.t0} \sum\nolimits_{i=1}^N\frac{\ttau_i\delta^2_i(\hmu,\hsig)}{\hsig+s^2_i}\leq\sum\nolimits_{i=1}^N\frac1{\hsig+s^2_i}.
		\end{IEEEeqnarray}
		
		On both sides of \refe{eqn:tme.Etau.t1}, divide by $\hsig+s^2_i$  and take the sum over $i$ from 1 to $N$, yielding
		\begin{IEEEeqnarray}{rCl}\label{eqn:tme.Etau.t2}
		\hskip-1.5em\sum\nolimits_{i=1}^N\frac{\hnu+1}{\hsig+s^2_i}=\sum\nolimits_{i=1}^N\frac{\ttau_i{\delta_i^{2}}(\hmu,\hsig)}{\hsig+s^2_i}+ \sum\nolimits_{i=1}^N\frac{\hnu\ttau_i}{\hsig+s^2_i}.
		\end{IEEEeqnarray}
		Substituting 
		\refe{eqn:tmeta.s2.t} and \refe{eqn:tmeta.s2.t0} into \refe{eqn:tme.Etau.t2}, respectively, we obtain, when $\hsig>0$,
		\begin{IEEEeqnarray}{rCl}\label{eqn:m.Etau.t1}
		\sum\nolimits_{i=1}^N\frac{\ttau_i}{\hsig+s^2_i}=\sum\nolimits_{i=1}^N\frac{1}{\hsig+s^2_i}, 
		\end{IEEEeqnarray}
		and when $\hsig=0$,
		\begin{IEEEeqnarray}{rCl}\label{eqn:m.Etau.t10}
		\sum\nolimits_{i=1}^N\frac{\ttau_i}{\hsig+s^2_i}\geq\sum\nolimits_{i=1}^N\frac{1}{\hsig+s^2_i}, 
		\end{IEEEeqnarray}
		
		When $\hsig>0$, from \refe{eqn:m.Etau.t1} we have
		\begin{IEEEeqnarray*}{rCl}\label{eqn:m.Etaua}
		\frac1N\sum\nolimits_{i=1}^Nu_i\ttau_i=1,
		\end{IEEEeqnarray*}
		and when $\hsig=0$, from \refe{eqn:m.Etau.t10} we have
		\begin{IEEEeqnarray*}{rCl}\label{eqn:m.Etaua0}
		\frac1N\sum\nolimits_{i=1}^Nu_i\ttau_i\geq1.
		\end{IEEEeqnarray*}
		where $u_i$ is given by \refe{eqn:ui}. This completes the proof.
	\end{proof}
	\end{appendices}
	\bibliography{journals_wiley,lit,jhzhao-pub}
	\bibliographystyle{elsarticle-harv}
	
\end{document}